# Improved theoretical prediction of nanoparticle sizes with the resistive-pulse technique

Zihao Gao,[1,2] Long Ma,[1,2] Zhe Liu,[1,2] Jun Huang,[1] Hanlian Liu,[1] Chuanzhen Huang,[1,3] and Yinghua Qiu[1,2]*

1. Key Laboratory of High Efficiency and Clean Mechanical Manufacture of Ministry of Education, National Demonstration Center for Experimental Mechanical Engineering Education, School of Mechanical Engineering, Shandong University, Jinan, 250061, China

2. Shenzhen Research Institute of Shandong University, Shenzhen, 518000, China

3. School of Mechanical Engineering, Yanshan University, Qinhuangdao, 066004, China

*Corresponding author: yinghua.qiu@sdu.edu.cn

**ABSTRACT**

With the resistive-pulse technique (RPT), nanopores serve as the nanofluidic sensors of various analytes for their many physical and chemical properties. Here, we focus on the size measurement and its theoretical prediction for sub-200 nm nanoparticles with RPT. Through systematical investigation of the current blockade of nanoparticles across cylindrical nanopores with simulations, Maxwell's method considering the shape coefficient and access resistances agrees well with simulation results. However, the widely used integration method of the resistance has distinct deviations in various cases. With the introduction of a correction factor $β$ to the integration method, our revised equations can provide good predictions for simulation results. $β$ shows a strong dependence on the diameter ratio ($d/D$) of the nanoparticle and nanopore. Following the same strategy, modified equations are provided for the accurate size prediction for nanoparticles across conical nanopores, where the integration method is the default convenient way. The correction factor $β′$ relates to $β$ in cylindrical nanopores. $β′$ exhibits independence on the pore geometry parameters and diameters of nanoparticles, but dependence on the surface charge density of conical nanopores. Our improved equations can provide theoretical predictions for the accurate size detection of 100-200 nm diameter nanoparticles across cylindrical and conical nanopores.

## I. INTRODUCTION

With the resistive-pulse technique, nanopores with various diameters provide a versatile platform for the nanofluidic sensing of different analytes,[1-3] such as biomolecules like DNA, RNA, and protein molecules,[4] nanoparticles like liposomes/exosomes,[5, 6] and artificial polystyrene balls,[7, 8] as well as cells.[9] The resistive-pulse technique, originating from the Coulter counter,[10] is a kind of high-throughput and non-contact sensing method under aqueous solutions, which can provide much information about physical and chemical properties of the analytes, such as concentration,[6] shapes,[8] sizes,[11, 12] and surface charges.[13, 14]

During the resistive-pulse detection, the ionic current through nanopores is recorded. The analyte can be moved through the nanopore via electrophoresis, electroosmotic flow, or hydrostatic pressure, which induces a transient change in the current trace under the applied voltage, called a resistive pulse.[2] When the pore is open, i.e. no analyte is present, the obtained current is $I_o$. During the translocation of an analyte, the blockade appears in the current trace. The blocked ionic current is $I_b$ which induces the magnitude of the pulse as $\Delta I = I_o - I_b$.[8] The width and number per unit time of these pulses correspond to the duration time and frequency. The amplitude and duration time of obtained pulses are the measure of the size and surface charge of the analyte.[13] The pulse frequency can shed light on the concentration of the analyte in the test sample.[15] From the details of the resistive pulses, the softness or deformation of the analytes can also be investigated and analyzed.[5, 16, 17]

Here, we focus on the size measurement of nanoparticles through nanopores based on the resistive-pulse technique. Compared with other methods for the detection of nanoparticles, the resistive-pulse technique can provide highly repeatable signals for the detection of individual particles, and reciprocating detection of the same particle many times by changing the polarity of the applied voltage.[8, 18]

Via micropores, the resistive-pulse technique was first used to detect the concentration and size distribution of cells in aqueous solutions.[19] Since the resolution of the analyte detection is determined by the magnitude of the current blockade, which is proportional to the volume ratio of the particle to the pore.[20-22] It is important to select nanopores with an appropriate size for the detection of specific particles to obtain a good sensitivity.[6] With the development of micro and nanofabrication techniques, pores with sub-micron diameters and lengths can be prepared conveniently,[1] which further expands the detection range of the resistive-pulse technique to small nanoparticles, such as exosomes/liposomes and viruses, whose size ranges from ~50 to ~200 nm.

With the recorded current traces, the accurate relationship between the current blockade and the particle size is essential for the theoretical prediction of the nanoparticle size.[23] According to Ohm's law, the classical theory describes the magnitude of the current blockade with the expression of the resistance change during the translocation of nanoparticles through the pore, which is generally applicable for particles larger than 200 nm in diameter.[8, 23] From the literature, the resistance change in the nanopore system can be described with Maxwell's approximation and the resistance integration method.[23]

DeBlois et al.[23] experimentally verified Maxwell's method and found that when the pore length was comparable to or larger than the pore diameter, the experimental result is not in agreement with Maxwell's method. They corrected the pore length for $L=D$ and added a correction factor for $L>D$ during theoretical calculations to obtain a good agreement, respectively. Grover et al.[24] and Hurley et al.[22] proposed a theory, which considered that the electric field applied in the nanopore is uniform, and the resistance change during the translocation of nanoparticles through the pore is proportional to its volume. DeBlois et al.[23] considered that the change of the electric field near the particles was not included in the resistance integration method. By solving the Laplace's equation, the potential distribution of the particles in the cylindrical pore was obtained, and a correction factor $F$ was proposed to correct the electric field change caused by the particles in the theoretical calculation. Willmott et al.[25] found that the end effect was not considered in the resistance integration method. When the particle is located at the tip boundary of the conical nanopore, they construct an artificial cone to approximately estimate the electric field around the particle.

From the literature, the quantitative relationship between the particle size and current blockade for sub-200 nm particles is rarely involved.[8, 23, 26] For the size prediction for nanoparticles, Maxwell's method has its limitations because it is derived under the assumption that $d<<D$ and suitable for cases with cylindrical pores.[23] The integration method can be employed in cases with pores of various shapes, such as cylindrical, and conical, which provides the accepted theoretical prediction for the size measurement with particles in conical nanopores.[27] For small particles, the resistance change obtained with the integration method is ~2/3 times that calculated using Maxwell's method.[28] The reason of this deviation involves the non-uniformity of the electric field in the nanopore, the non-uniformity of ions concentration in the solution and the complex influencing factors such as electroosmotic flow.[23, 28, 29] To achieve accurate prediction of particle sizes, the empirical correction factor of 3/2 is usually added in the integration method, which may cause deviations in cases involving alterations in $d/D$ or pore length, as well as the incorporation of factors such as access resistance and shape coefficient.[21]

In this work, systematical simulations have been conducted to investigate the current blockade generated during the translocation of sub-200 nm diameter nanoparticles through nanopores. Then, both Maxwell's method and integration method were used to predict the obtained current blockade theoretically. We found that Maxwell's method can provide a perfect fit for the current blockade of sub-200 nm diameter spheres in cylindrical nanopores which should include the access resistance and shape coefficient. For the integration method, an extra correction factor $β$ is essential for the accurate prediction of the simulation results in cylindrical nanopores, which depends strongly on the ratio of $d/D$. For different selections of $d$ and $D$ but equal $d/D$, the correction factor $\beta$ is the same which is applicable for $d/D$ varying from 0.2 to 0.8. Following the same strategy, a correction factor $\beta'$ is introduced in the integration method for the theoretical prediction of the current blockade in conical nanopores. This correction factor $\beta'$ has a certain relationship with that in cylindrical nanopores, which is independent of the pore geometry parameters and particle sizes but correlated with the surface charge density of nanopores. With our revised equations, the accurate theoretical prediction for particle sizes can be achieved in resistive-pulse detection with conical pores.

**II. METHODOLOGY**

Maxwell's approximation[30] considers the influence of insulating nanoparticles on the resistivity of the solution. The resistance change caused by a nanoparticle staying in the nanopore can be described with Eq.1.[23] This equation is denoted as Maxwell's method here.

$$\Delta R = R'_p - R_p = \frac{4\rho d^3}{\pi D^4} \tag{1}$$

where $d$ and $D$ are the particle diameter and pore diameter, respectively. $R'_p$ and $R_p$ are the pore resistances with and without the particle inside the nanopore. $\rho$ is the resistivity of bulk solutions.

Assuming a uniform resistivity in the whole system, the resistance inside the nanopore $R_p$ can be roughly evaluated by the integration method along the pore length. For a unit length $dL$, the resistance ($dR$) of a cylindrical unit can be expressed by the ratio of the length and cross-sectional area ($A_1$) analogous to a solid wire.[23, 31]

$$dR = \frac{\rho \cdot dL}{A_1} = \frac{4\rho \cdot dL}{\pi D^2} \tag{2}$$

$$R_p = \rho \int_0^L \frac{dz}{A_1(z)} = \frac{4\rho L}{\pi D^2} \tag{3}$$

where $L$ and $A_1(z)$ are pore length and the local cross-sectional areas of the nanopore.

Then, the resistance change caused by the nanoparticle of diameter $d$ inside nanopores with complex geometry can be calculated by the integration of unit resistance along the pore length (Eq.4).[11, 27] We refer to Eq.4 as the resistance integration method.

$$\Delta R = \rho \left[ \int_0^d \frac{dz}{A_1(z) - A_2(z)} - \int_0^d \frac{dz}{A_1(z)} \right] \quad (4)$$

in which $A_2(z)$ is the local cross-sectional area of the nanoparticle at $z$ along the pore axis.

With the resistance integration method, Gregg et al.[28] obtained the analytical equation of resistance change caused by the particle in the cylindrical pore.

$$\Delta R = \frac{4\rho}{\pi D} \left[ \frac{\arctan\left[ \frac{2(d/D)\sqrt{1-(d/D)^2}}{1-2(d/D)^2} \right]}{2\sqrt{1-(d/D)^2}} - \frac{d}{D} \right] \quad (5)$$

Inspired by the analysis of the flow around a circular tube by Smythe[32] and Cooke,[33] a shape coefficient $S(d/D)$ was proposed in the change of resistance when the particle appears in the pore.

$$S(d/D) = \frac{1}{1 - 0.8(d/D)^3} \quad (6)$$

Please note: the accurate theoretical prediction for the nanopore resistance requires the consideration of the access resistance ($R_{ac}$) in the system.[34] For micro and nanopores with a large length-diameter ratio, the contribution of the access resistance to the total resistance is negligible. While the access resistance in the pores with a small aspect ratio may account for more than 50% of the total

resistance.[31]

2D axisymmetric simulation models were built with COMSOL Multiphysics.[11, 35-37] Coupled Poisson-Nernst-Planck equations and Navier-Stokes equations were solved in our simulation models to consider the ionic distribution near the charged surface, ionic transport, and fluid movement in aqueous solutions. As shown in Figure 1, the nanopore is located between two reservoirs of 5 μm in diameter and length. The nanoparticle moved along the pore axis to mimic the translocation across the nanopore. In this work, sub-200 nm nanoparticles were focused on which diameter ($d$) changes from 100 to 200 nm, which is similar to the characteristic sizes of exosomes.[6] Due to the uniform diameter in the axial direction, cylindrical nanopores are usually fabricated on thin SiN membranes for particle detection.[5, 26, 38-40] The diameter of the nanopore ($D$) was set from 125 to 1000 nm to consider the variation of $d/D$ from 0.2 to 0.8, correspondingly. The pore length ($L$) varied from 150 nm to 2000 nm to consider the nanopores with different length scales. The surface charge density of cylindrical nanopores ($\sigma$) was mainly set as −0.005 C/m$^2$.[35, 41]

Conical nanopores were also considered in this work to mimic the cases with polymer nanopores.[11] We have considered the influence of the tip diameter ($D_{tip}$), pore length ($L$), and half-cone angle on particle detection. The half-cone angle ($\alpha$) was considered as 1.9º, 2.5º, and 3º.[11] To keep the same $d/D_{tip}$ ratio, the tip diameter used was 292 nm. The pore length varied from 8 to 14 μm. The default tip diameter, pore length, particle diameter, and half-cone angle were 292 nm, 11 μm, 200 nm, and 1.9°, respectively. The surface charge density of the pore ($\sigma$) was set from −0.005 to −0.08 C/m$^2$,[11] with a default value of −0.005 C/m$^2$.

In the simulations, the surface charge density of the nanoparticle was varied from −0.004 to −0.08 C/m$^2$ with the default value of −0.004 C/m$^2$.[5, 42] 0.1 M KCl solution was filled in the system as the default electrolyte for most cases. Other salt types, NaCl and LiCl, and concentrations of 0.01 and 0.05 M were also used to consider the influences of the ion diffusion coefficient and ionic concentration on current, respectively. The diffusion coefficients of $K^+$, $Na^+$, $Li^+$ and $Cl^-$ ions were $1.92\times10^{-9}$, $1.33\times10^{-9}$, $1.03\times10^{-9}$ and $2.03\times10^{-9}$ m$^2$/s.[11] The dielectric constant of water was set as 80. All simulations were conducted under the room temperature of 298 K with an applied voltage varying from 0.1 to 0.5 V with the default value of 0.5 V across the nanopores. To systematically investigate the change of blockade current caused by particles with different diameters across the pore, and include the influence of electric double layers near charged surfaces on ionic transport, the mesh size of 0.1 nm was used on the inner-pore surface and the particle surface. 0.5 nm mesh was chosen for the exterior membrane surfaces to lower the memory cost.[43-45] Detailed boundary conditions and the mesh size independence test were provided in Table S1 and Figure S1.[11, 35, 43-45]

The particles can have off-axis translocations through nanopores which may cause lager blockade current than along the axis of the pore under the applied voltage and pressure. The off-axis effect was first studied by Smythe et al[46], who provided a corresponding numerical relationship for the off-axis effect of spherical particles with different sizes. Berge et al.[47] found that the blockade current amplitude increases by less than 10% when the particle is displaced from the pore axis to the wall. Qin et al.[48] experimentally verified Berge's work and found that when the particles were large (such as $d/D$ =0.7), the results predicted by Berge's work had a large deviation. Weatherall et al.[49] considered the relationship of the off-axis particle trajectories and the blockade current amplitude. It can be found that the off-axis effect has little effect on the current blockade ratio $\Delta I/I_o$ before the particle approach the nanopore wall. For nanopores with a short length, the off-axis effect of the particles has a greater influence on the blockade current. Tsutsui et al.[50] found off-axis particles produce pulses up to 35% larger than those at the pore center in SiN$_x$ nanopores with small $L/D$. It is worth noting that we focus on the accurate theoretical prediction of particle sizes in resistive-pulse sensing with cylindrical and conical nanopores. Therefore, the influence of the off-axis[51-53] was not taken into consideration in our simulations with long nanopores.

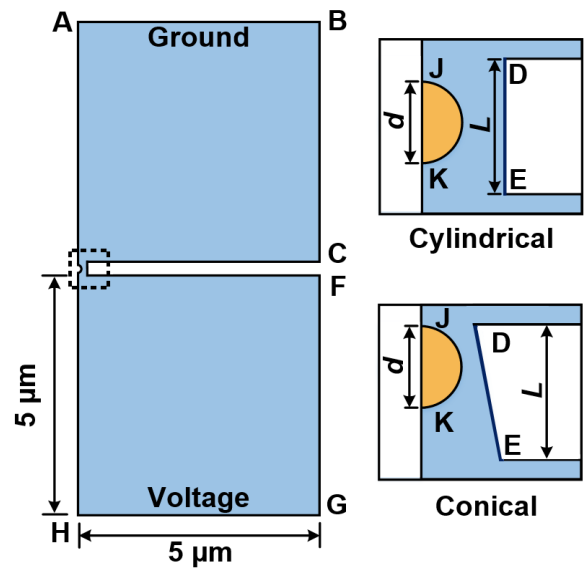

Figure 1 Scheme of nanofluidic simulations. The nanopore is located at the center, between two reservoirs with 5 μm in diameter and length. The nanoparticle with the diameter of *d* moves along the pore axis. The pore length is denoted as *L*. Zoomed-in parts on the right show the regions of nanopores with cylindrical and conical shapes.

### III. RESULT AND DISCUSSIONS

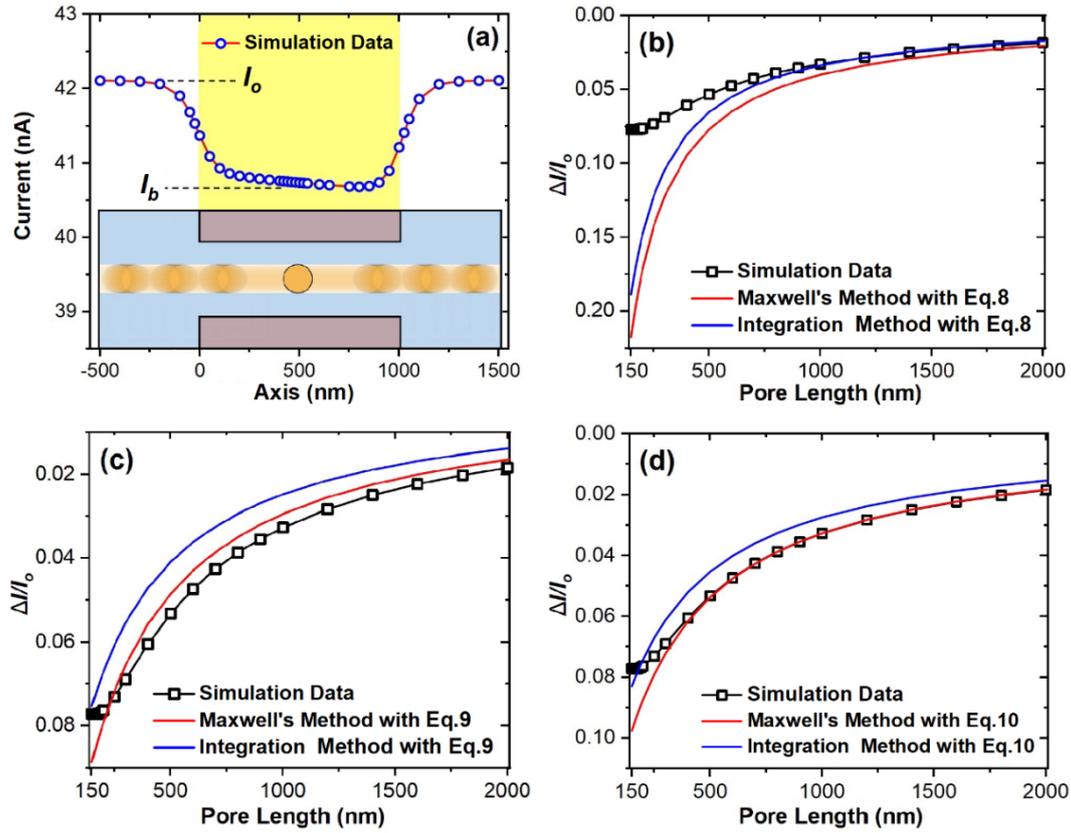

Figure 2 Current blockade ratio of a 150 nm diameter nanoparticle in 300 nm diameter nanopores with various lengths. (a) An example of the simulated current trace. $I_o$ and $I_b$ are the open current and blockade current, respectively. (b)-(d) The current blockade ratios obtained from simulations, accompanied by the corresponding theoretical predictions using Eq. 8 (b), Eq. 9 (c), and Eq. 10 (d).

With nanofluidic simulations, the current trace during the translocation of the nanoparticle through the nanopore can be obtained, which can be used for the measurement of the particle size based on the analysis of the current blockade ratio.[8] Using cylindrical nanopores, the current blockade was studied through a series of simulations with various values of *d/D* (Figure S2).

Figure 2a shows a simulated current trace of nanoparticle translocation through the nanopore, referred to as an event. Different current values are obtained by changing the position of nanoparticles along the pore axis. When the nanoparticle is located far away from the pore entrance, the ionic transport inside the pore is not influenced by the particle, which generates the open pore current $I_o$. As the nanoparticle approaches the pore entrance and travels through the nanopore, it inhibits the ionic transport inside the pore due to the physical blockade resulting in the blockade current $I_b$.[8] After the nanoparticle exits the nanopore, the ionic current recovers to the open pore current. From Figure 2, for the nanoparticle with a diameter of 150 nm, the current blockade ratios are obtained when it passes through 300 nm diameter nanopores with the length varying from 150 to 2000 nm. The current blockade ratio depends on the volume ratio between the nanoparticle and the nanopore. Though the nanoparticle and the pore keep the constant diameters, shorter nanopores produce higher current blockade ratios which means a higher detection precision in the resistive-pulse measurement.[26]

According to the ohmic relationship between the resistance and the current in the nanopore under applied voltages, the current blockade ratio $\Delta I/I_o$ can be expressed as the change in the resistance $\Delta R$ over the system resistance with the nanoparticle inside the pore $R_b$, Eq. 7.

$$\frac{\Delta I}{I_o} = \frac{I_o - I_b}{I_o} = \frac{R_b - R_o}{R_b} = \frac{\Delta R}{R_o + \Delta R} \tag{7}$$

where the resistance of the open pore $R_o = R_p + 2R_{ac} = \frac{4\rho L}{\pi D^2} + \frac{\rho}{D}$,[31, 54] and the resistance of the blocked nanopore $R_b = R_o + \Delta R$.

With the mathematical description of the nanopore resistance by Maxwell's approximation and integration method,[23] the magnitude of the current blockade can be predicted theoretically. As shown in Figures 2b-2d, the simulated results are theoretically fitted with both methods. Different equations are used to consider the influence of the shape coefficient[23] (Eq.8), the access resistance of the nanopore (Eq.9), or both (Eq.10). Note that the access resistance of nanopores was not considered in Maxwell's method, because of the later development of the theoretical description of the access resistance.[34] From our results, the shape coefficient is essential for the theoretical prediction which increases the resistance change.

$$\Delta I / I_o = \frac{\Delta R \cdot S(d/D)}{\Delta R \cdot S(d/D) + R_p} \tag{8}$$

$$\Delta I / I_o = \frac{\Delta R}{\Delta R + R_p + 2R_{ac}} \tag{9}$$

$$\Delta I / I_o = \frac{\Delta R \cdot S(d/D)}{\Delta R \cdot S(d/D) + R_p + 2R_{ac}} \tag{10}$$

The access resistance contributes an important portion of the total resistance, especially for short nanopores.[31] Figure 2c shows the theoretical prediction of the current blockade with both methods considering the access resistance with Eq. 9. When the pore length is longer than 200 nm, the deviation can be ignored without consideration of the access resistance. As the pore length gets shorter than 200 nm, the neglect of the access resistance in the theoretical prediction induces much larger deviations. In the theoretical prediction of the current blockade ratio obtained in short nanopores, the access resistance must be considered.

From Figure 2d, with the consideration of both access resistance and shape coefficient (Eq.10),[55] Maxwell's method can provide a good prediction of the current blockade ratio of nanoparticles in cylindrical nanopores. For cases with different sizes of nanopores and particles, as well as pore lengths, the prediction with Maxwell's method agrees well with the simulation results (Figure S2). Therefore, the mathematical fitting of the current blockade of weakly charged surface particles through cylindrical nanopores can be achieved by Maxwell's method considering both the access resistance and the shape coefficient, i.e. Eq.10.

However, the integration method[23, 27, 28, 56] which is widely accepted under various pore shapes has a large deviation from the simulation results. Please note that the case with *d/D*=0.5 shown here is used to exhibit the deviation between the theoretical prediction with the integration method and the simulation results, which will be further discussed later. Only when $d/D=\sim 0.6$, can the integration method agree well with the simulation. Therefore, it is significantly important to improve Eq.10 to provide an accurate prediction of the current blockade with the integration method.

$$\Delta I / I_o = \frac{\Delta R \cdot S(d/D) \cdot \beta}{\Delta R \cdot S(d/D) \cdot \beta + R_o} \quad (11)$$

By adding a correction factor, $\beta$, to Eq.10 to modulate the resistance variation induced by the particle translocation, the modified integration method (Eq.11) can also provide a good prediction for the simulation results like Maxwell's method. As shown in Figure 3, for the case of $d$=150 nm and $D$=300 nm, with a constant $\beta$=1.2, the integration method agrees well with the simulated blockage current for the particle translocation under various pore lengths(Figure S3). Note that $\beta$ is not a constant equal to 1.5,[23, 28] which depends on the ratio of $d/D$, instead of the individual magnitudes of $d$ or $D$. Due to the weak dependence of $\beta$ on the surface charge density of nanopores (see below), we use $\beta_{-0.005}$ to denote the $\beta$ values at $\sigma$=−0.005 C/m$^2$. Figure 3b illustrates the $\beta_{-0.005}$ values obtained from the blockage currents of nanoparticles with different diameters at $d/D$=0.5. Under a constant $d/D$, $\beta_{-0.005}$ remains unchanged for different particle diameters which facilitates the theoretical prediction of the current blockade ratio $\Delta I/I_o$.

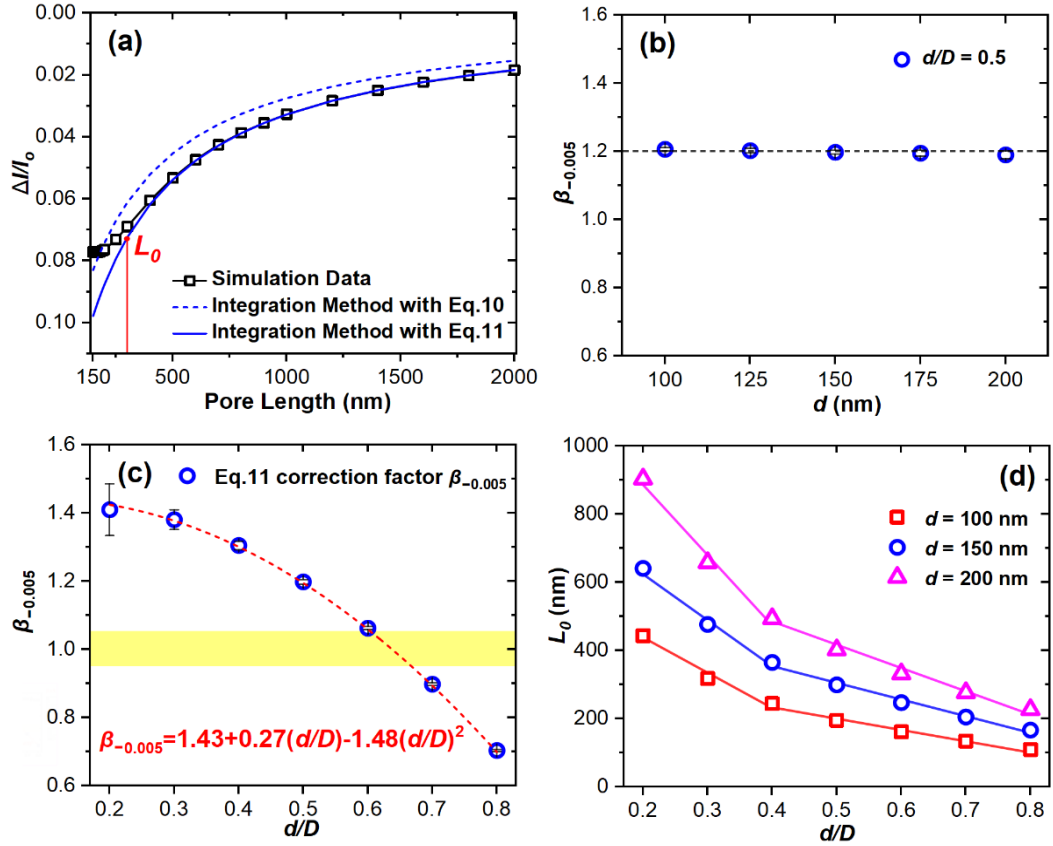

Figure 3 Theoretical prediction for the simulation results with the integration method containing a correction factor $\beta_{-0.005}$. $\beta_{-0.005}$ denotes the $\beta$ values in Eq. 11 at $\sigma=-0.005$ C/m². (a) Theoretical prediction for the simulation data with Eq.10 and Eq.11. The diameters of the nanoparticle and the nanopore are 150 and 300 nm, respectively. The solid red line shows the pore length ($L_0$) at which the deviation of 5% appears between theoretical prediction with Eq.11 and simulation results. (b) Values of $\beta_{-0.005}$ for different particle sizes under $d/D$=0.5. Each point was the average over those with various pore lengths from 500 to 3000 nm. (c) Values of $\beta_{-0.005}$ for different $d/D$. The red-dash line is the theoretical description of the correction factors. Each point was averaged over those numbers obtained in the cases with the particle diameter varying from 100 to 200 nm, and the pore length changing from 500 to 3000 nm. (d) The pore length $L_0$ obtained under various values of $d/D$ with particle diameters changing from 100 to 200 nm.

Systematic simulations were conducted with various $d/D$ by adjusting $d$ from 100

to 200 nm and setting $D$ according to the ratio of $d/D$ changing from 0.2 to 0.8. For each selected $d$ and $D$, the pore length varies from 150 to 3000 nm. The derived $\beta$ values are depicted in Figure 3c, which exhibits a direct correlation with $d/D$. For each point, the error bar represents the standard deviation among the values of $\beta$ with the same $d/D$, but different diameters of the nanoparticle and nanopore, and pore lengths. With $d/D$ changing from 0.2 to 0.8, $\beta_{-0.005}$ decreases from 1.4 to 0.7. Please note when $d/D$ approaches ~0.6, the correction factor $\beta_{-0.005}$ is ~1, which can be neglected in Eq.11. Based on the dependence of $\beta$ on $d/D$, the numerical fitting of the $\beta_{-0.005}$ values is given as shown by the red-dash line in Figure 3c. Note that $\beta_{-0.005}$ has a variation from 1.35 to 1.5 at a small ratio of $d/D$=0.2. Then, the application of the integration method containing the correction factor $\beta$ may cause a deviation in the theoretical prediction when detecting small particles using wide nanopores. This emphasizes the importance of selecting appropriate nanopores for specific analytes.

From Figure 3a, the revised equation with a correction factor (Eq.11) can fit the blockade current in most cases, especially with long nanopores. However, significant deviations between the theoretical prediction with Eq. 11 and simulation results appear when the nanopore is short. We think this may be attributed to the influence of particles on the access resistance.[57] In cases with short nanopores, the access resistance ($R_{ac}$) dominates the total resistance, which can be affected by the appearance of the nanoparticle inside the nanopore. To evaluate the accuracy of the modified equation in the prediction of the blockade currents, a deviation within 5% is defined as a good agreement between the theoretical predictions and the simulation results. The pore length $L_0$ corresponding to a 5% deviation serves as the lower threshold for the applicability of the modified equation, depicted as the solid red line in Figure 3a.

Figure 3d shows the values of $L_0$ obtained from the theoretical prediction with Eq. 11 to the blockade current of nanoparticles 100, 150, and 200 nm in diameter. As $d/D$ increases from 0.2 to 0.8, $L_0$ gradually decreases from ~900 nm to ~100 nm. With a constant $d/D$, $L_0$ exhibits a linear increase with $d$ or $D$. At large $d/D$ ratios, $L_0$ has a small value which shows a weak increase with the particle diameter. With the variation of $L_0$ under different conditions, an appropriate pore length should be selected for particle detection if the rough sizes of the particle and the pore are known. For the nanopores longer than $L_0$, the developed equations here can provide accurate theoretical predictions for the obtained current blockades.

In resistive-pulse detections, both cylindrical and conical nanopores have been widely used. The blockade current amplitude of nanoparticles through cylindrical nanopores can be theoretically predicted using the aforementioned Maxwell's method and integration method. We have shown that the current blockades in cylindrical nanopores can be well predicted by Maxwell's method with the consideration of the access resistance $R_{ac}$ and the shape coefficient $S(d/D)$ (Eq.10), and the integration method containing the correction factor (Eq.11). Conical nanopores can be prepared with various manufacturing techniques, such as drilling with the focused ion beam,[58] track etching,[11] pulling of fused glass pipettes,[59, 60] and puncturing with tungsten needles.[61] During the resistive-pulse detection with long nanopores, conical nanopores can provide a much better resolution than cylindrical ones due to their asymmetric pore shape.[6] When the particle stays at the pore tip, a distinct blockade current is obtained because of the dominated resistance of the tip region in the total resistance, which facilitates the detection of small nanoparticles.[21] While for the resistive-pulse detection with conical nanopores, the integration method provides the only convenient way for the theoretical prediction of the current blockade (Eq.12).[11, 27, 28]

$$\Delta R = \frac{\rho}{\pi\sqrt{A}\sqrt{C-\frac{B^2}{4A}}}\left[\arctan\left(\frac{\sqrt{A}d+\frac{B}{2\sqrt{A}}}{\sqrt{C-\frac{B^2}{4A}}}\right) - \arctan\left(\frac{\frac{B}{2\sqrt{A}}}{\sqrt{C-\frac{B^2}{4A}}}\right)\right] \quad (12)$$

where,

$$A = \left(\frac{D_{base} - D_{tip}}{2L}\right)^2 + 1,$$

$$B = 2x_0 \left(\frac{D_{base} - D_{tip}}{2L}\right)^2 + D_{tip}\left(\frac{D_{base} - D_{tip}}{2L}\right) - d,$$

$$C = x_0^2 \left(\frac{D_{base} - D_{tip}}{2L}\right)^2 + x_0 D_{tip}\left(\frac{D_{base} - D_{tip}}{2L}\right) + \left(\frac{D_{tip}}{2}\right)^2,$$

$x_0$ is the position of the particle edge from the nanopore tip. $D_{base}$ and $D_{tip}$ are the diameters of the large and small openings of the conical nanopore.

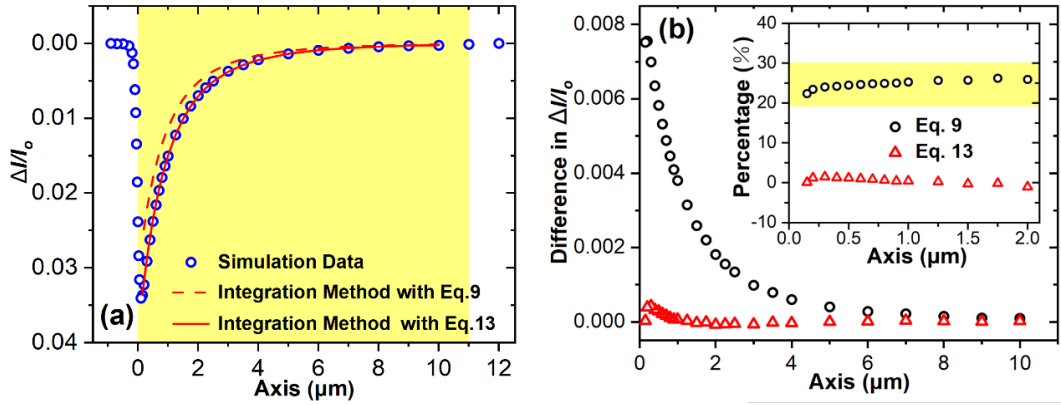

Figure 4 Theoretical prediction for the simulation results in conical nanopores with the integration method containing a correction factor $\beta'$. (a) Simulated current blockade ratios and the theoretical predictions with Eq.9, and Eq.13. The tip diameter, pore length, particle diameter, and half-cone angle were 292 nm, 11 μm, 200 nm, and 1.9° respectively. (b) The deviation between the simulated current blockade ratio and the theoretical predictions with Eq.9, and Eq. 13. The inset shows the percentage of deviation.

Figure 4a shows the current blockade during the translocation of a 200 nm diameter particle through a conical nanopore with 292 and 1022 nm in the tip and base openings, and 11 μm in length.[11] When the particle stays near the base, small magnitudes of the current blockade are obtained due to the wide local region of the nanopore. As the particle moves toward the tip opening, significant blockades on the ion transport through the nanopore can induce large magnitudes of the current blockade. Eq.9 is used to fit the simulation results. While similar to our previous study,[11] a deviation of more than 20% near the pore tip appears between the predicted values and simulations. This is unfavorable for the precise measurement of particle sizes using conical nanopores. In practical nanofluidic experiments, the maximum current blockade is the important characteristic in the events which is usually for the size prediction of nanoparticles.

Following a similar strategy for the prediction of $\Delta I/I_o$ in cylindrical nanopores, modification is made in Eq.9 by adding a shape coefficient $S(d/D)$ and a correction coefficient $\beta'$ for conical nanopores, resulting in Eq.13. Note that in conical nanopores, $D$ represents the local pore diameter at the center of the nanoparticle. Due to the varying diameter along the axis of the conical nanopore, $\beta'$ is not a constant value that depends on the axial position of the particle. According to the ratio of $d/D$, and the values of the correction coefficient $\beta$ in cylindrical nanopores, Eq.13 enables the theoretical prediction of the blockade current of nanoparticles in the conical pore. For our case in Figure 4a, when the particle is located within 2 μm away from the tip boundary, a deviation less than 1.5% is obtained between the theoretical prediction with the correction factor $\beta'$ and the simulation results.

$$\Delta I/I_o = \frac{\Delta R \cdot S(d/D) \cdot \beta'}{\Delta R \cdot S(d/D) \beta' + R_o} \qquad (13)$$

where

$$D = D_{tip} + \left(x_0 + \frac{d}{2}\right) \cdot \tan\alpha$$

$$R_o = R_p + R_{ac} = \frac{4\rho L}{\pi D_{tip} D_{base}} + \frac{2\rho}{D_{tip} + D_{base}}$$

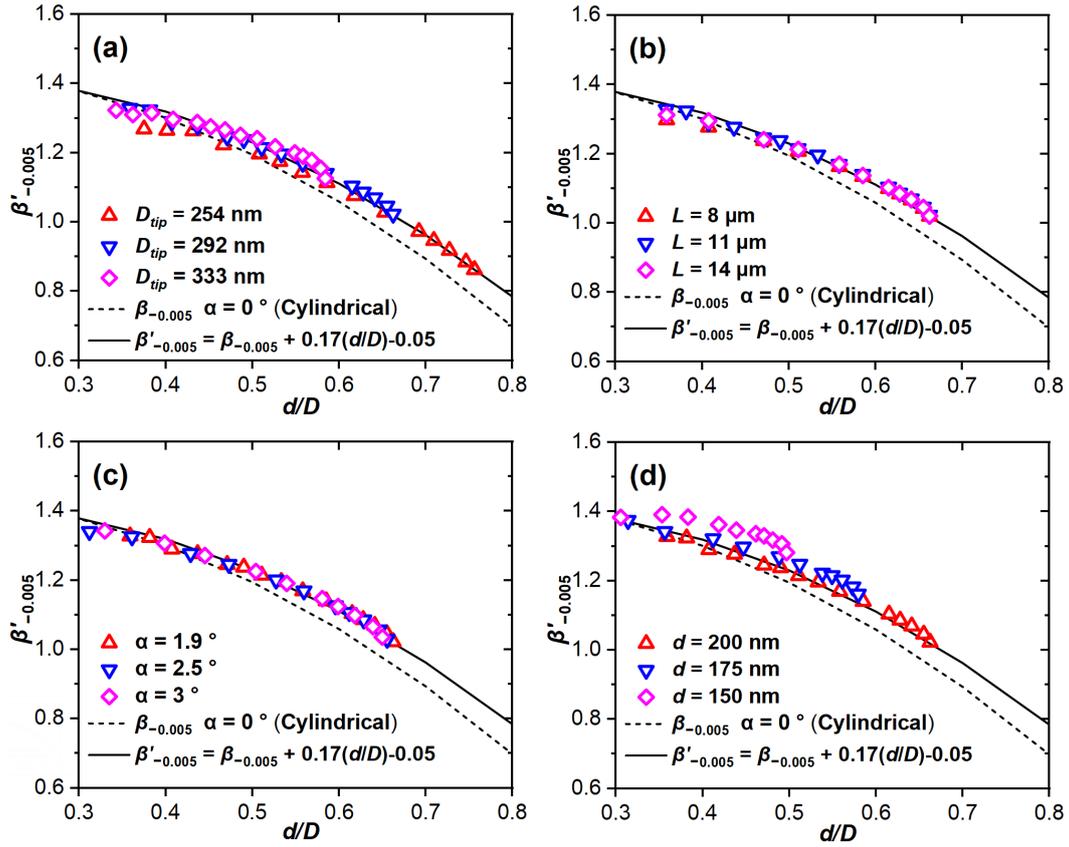

Figure 5 Values of $\beta'_{-0.005}$ under various conical nanopore parameters and particle diameters. $\beta'_{-0.005}$ denotes the $\beta'$ values in Eq. 13 at σ=−0.005 C/m². The default tip diameter, pore length, particle diameter, and half-cone angle were 292 nm, 11 μm, 200 nm, and 1.9°, respectively. Dash lines depict the values of $β$ for cylindrical nanopores at σ=−0.005 C/m² under various d/D. Solid lines correspond to the fitting of the correction factor $\beta'_{-0.005}$. (a) Tip diameters of 254, 292, and 333 nm. (b) Pore lengths of 8, 11, and 14 μm. (c) Half-cone angles of 1.9°, 2.5°, and 3°. (d) Particle diameters of 200, 175, and 150 nm. Please note that in conical nanopores, D means the local pore diameter at the center of the nanoparticle.

Further simulations have been conducted to investigate the influence of the pore parameters, including the geometry parameters and surface charge density, as well as the particle diameter on the correction factor $β'$. Figure 5 shows the variation of the correction factor $β'$ to d/D under different tip diameters $D_{tip}$, pore lengths L, half-cone angles α, and particle diameters d. Due to the relatively high detection accuracy in the narrow regions for particle detection,[11, 21] we mainly focus on the blockade current obtained within the 6 μm near the tip opening. The results reveal that $\beta'$ is almost independent of the geometry parameters of conical nanopores and particle diameters. The correction factor for conical nanopores can be obtained from those in cylindrical pores by $\beta'_{-0.005}$ = $\beta_{-0.005}$+0.17(*d/D*)−0.05. Here, $\beta'_{-0.005}$ and $\beta_{-0.005}$ are used to denote the $\beta'$ and $\beta$ values at σ=−0.005 C/m$^2$, respectively. For our simulations with differently sized nanoparticles and various applied conditions such as voltage, salt concentration, and electrolyte type, the maximum blockade current induced by the particle translocation can be roughly obtained when the edge of nanoparticles contacts the tip opening with the known geometry parameters of conical nanopores (Figure S4 and Figure S5). Inside the nanopore within ~200 nm away from the tip opening, the obtained blockade current of a small nanoparticle shares a similar value. Since it is difficult to determine the axial positions of particles inside conical nanopores from the event, the maximum blockade current is crucial due to its uniqueness. By setting the corresponding value of *D*, the prediction of the maximum blockade current generated by nanoparticles near the tip opening can be achieved.

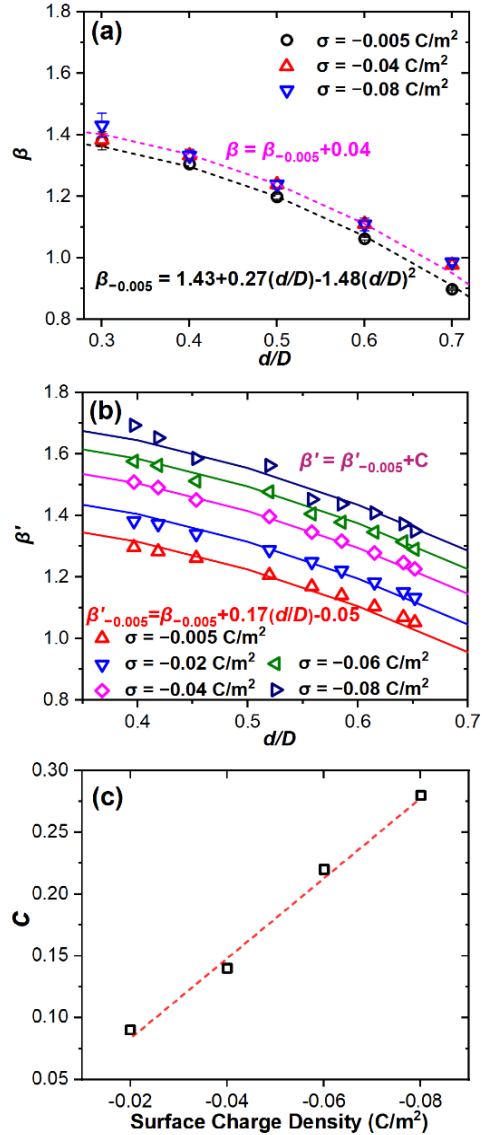

Figure 6 Dependence of $\beta$ and $\beta'$ on the surface charge density of the nanopore. (a) The variation of $\beta$ on $d/D$ under different surface charge densities. The black-dash line depicts the theoretical fitting of $\beta_{-0.005}$ in cylindrical nanopores. The pink-dash line corresponds to the fitting of $\beta$ under various surface charge densities of cylindrical nanopores. (b) The variation of $\beta'$ on $d/D$ under different surface charge densities. The black-dash line depicts the theoretical fitting of $\beta_{-0.005}$ in cylindrical nanopores. Solid lines correspond to the fitting of $\beta'$ under various surface charge densities of the conical nanopores. (c) The off-shift for the fitting of $\beta'$ under different surface charge densities from that in cylindrical nanopores. The tip diameter, pore length, particle diameter, and half-cone angle were 292 nm, 11 μm, 200 nm, and 1.9°,

respectively.

The surface charge density is another important parameter of nanopores[62] that can modulate the ion concentration and fluid flow inside the nanopore.[11, 29] Based on the obtained blockade current ratios in cylindrical nanopores, we explored the influence of the surface charge density of nanopores on the fitting parameter $\beta$. Here the d/D ratios from 0.3 to 0.7 were focused on considering the practical situation in nanofluidic experiments. As exhibited in Figure 6a, the $\beta$ values with d/D ratios under different charge strengths share a similar profile. Parallel curves with an offset to the one under σ=−0.005 C/m$^2$ can be used to fit those under other surface charge densities through $\beta=\beta_{-0.005} + C$, in which $C$ is a constant. With the surface charge density of pore walls increasing from −0.005 to −0.08 C/m$^2$, $C$ has a small value of ~0.04 which weakly depends on the surface charge density.

Referring to our previous work with a charged conical nanopore under −0.04 C/m$^2$,[11] the blockade current $\Delta I/I_o$ can be well predicted using Eq.13 with a correction factor $\beta'_{-0.04} = \beta'_{-0.005} + 0.09$ (Figure S6). This shows that the value of the correction factor can also be influenced by the surface charge density in conical nanopores. For conical nanopores, the correction factor $\beta'$ has been obtained under various surface charge densities from −0.005 to −0.08 C/m$^2$, as shown in Figure 6c. In nanopores with a higher surface charge density, a larger $\beta'$ is required due to the modulation of surface charges in the current blockade (Figure S7). Under each surface charge density, the obtained values of $\beta'$ share a similar trend to that in cylindrical nanopores. Parallel curves are also used to fit the values of $\beta'$ under different surface charge densities through $\beta' = \beta'_{-0.005} + C$. From Figure 6b, $\beta'$ is approximately linearly correlated with surface charge density. This may provide a valuable reference for the selection of correction factors in particle size prediction with differently charged conical nanopores. Similarly, the surface charge densities of the particles can affect the current blockade (Figure S8) due to the modulation in ionic concentration inside the nanopore during the translocation of nanoparticles. Please note that we haven't attempted to provide an accurate theoretical prediction for all situations. In our equations, the correction factor can be affected by the surface charges of nanopores and nanoparticles, as well as the salt concentration (Figure S5 and Figure S8). Before the application of our equations under some other specific conditions, the correction factor $\beta'$ needs to be determined.

Our developed equations were then used to predict the particle size based on the experimental data in our earlier research. In Ref.[11], the neutral polystyrene particles with 280 nm in diameter were detected via the RPT with a conical polymer micropore. Figure S9 shows the comparison among current blockade ratios obtained from the nanofluidic experiment and theoretical predictions with the conical nanopore. Eq. 9 cannot be applied directly to predict the blockade current, which has a large deviation of ~20% from the experiment data. This is unfavorable for the precise measurement of particle sizes using conical nanopores. With Eq. 13 containing the correction factor $\beta'$=1.38, we achieved a good theoretical prediction of the current blockade value in the experiment.

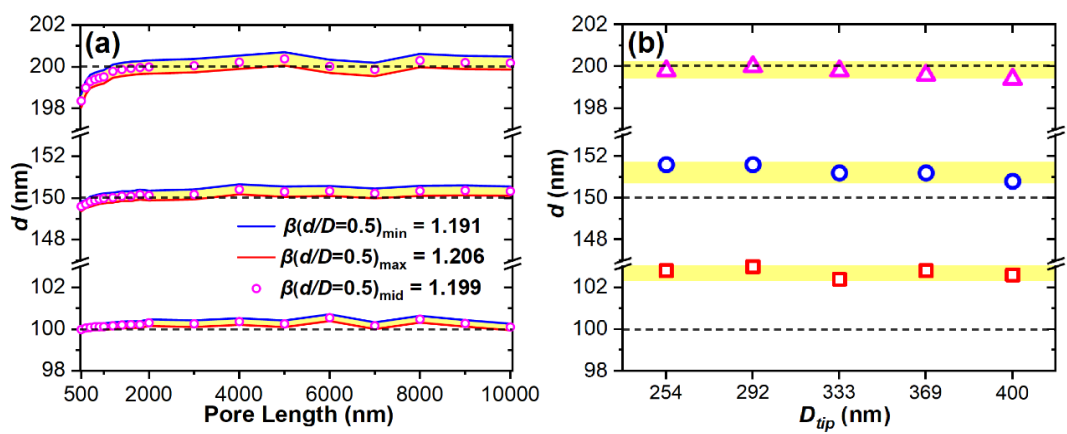

Figure 7 Predicted size with the revised equations containing correction factors. (a) Predicted particle diameter for nanoparticles of 100, 150, and 200 nm in diameter using the integration method containing the correction factor with cylindrical nanopores under various pore lengths from 500 to 10000 nm. Please note that the minimum, median, and maximum values of $\beta$ are derived from the deviation of $\beta$ in Figure 3c. (b) Predicted particle diameter for nanoparticles of 100, 150, and 200 nm in diameter using the integration method containing the correction factor applied to conical nanopores with different tip diameters. The pore length, particle diameter, and half-cone angle were 11 μm, 200 nm, and 1.9º, respectively.

Finally, the accuracy of the size prediction is evaluated by the revised equations with correction factors as shown in Figure 7. Based on the fitting equation for cylindrical nanopores in Figure 3c $\beta_{-0.005}=1.43+0.27(d/D)-1.48(d/D)^2$, the particle size can be predicted from the measured current blockade due to the only unknown parameter is $d$ if the rough sizes of the pore are known. Considering the standard deviation of $\beta$, the predicted particle size has a sub-1% deviation from its physical size of 100~200 nm (Figure 7a). For conical nanopores, following the same strategy, with the fitting equation of the correction factor $\beta'_{-0.005}= \beta_{-0.005}+0.17(d/D)-0.05$ and Eq.13, the predicted size in conical nanopores with different tip diameters has a maximum deviation of ~3% from the real diameter of 100~200 nm (Figure 7b).

## IV. CONCLUSIONS

The resistive-pulse detection of sub-200 nm nanoparticles has been conducted through systematic simulations with cylindrical and conical nanopores. The obtained current blockade during the translocation of nanoparticles through nanopores has been theoretically predicted by Maxwell's method and the integration method. To agree well with simulation results in cases of cylindrical nanopores, the shape coefficient, and access resistance are required to be considered in Maxwell's method. For the widely accepted integration method, a correction factor $\beta$ is introduced to the equation to eliminate the deviation between simulation results and theoretical predictions, which only depends on the ratio of $d/D$. However, the influence of particles inside short nanopores on the access resistance induces the deviation between theoretical predictions and simulations, which limits the application of the revised equation. The minimum applicable pore length $L_0$ for theoretical predictions can provide a reference for the selection of pore length in resistive-pulse detection.

Modified equations for current blockade through conical nanopores are provided by introducing another correction factor $\beta'$, which correlates to $\beta$ in cylindrical nanopores. $\beta'$ depends only on the surface charge density on pore walls instead of the geometry parameters of conical nanopores and particle diameters. Our modified equations can provide good predictions of the current blockade in both cylindrical and conical nanopores, which are of significant importance for the accurate prediction of the particle diameters with the resistive-pulse technique.

## SUPPLEMENTARY MATERIAL

See supplementary material for additional simulation details, simulation results, and theoretical predictions.

## ACKNOWLEDGMENTS

This research was financially supported by the National Natural Science Foundation of China (52105579), the Natural Science Foundation of Shandong Province (ZR2020QE188), the Guangdong Basic and Applied Basic Research Foundation (2019A1515110478), and the Qilu Talented Young Scholar Program of Shandong University.

## AUTHOR DECLARATIONS

**Conflict of Interest**

The authors declare no competing financial interest.

**Author Contributions**

Zihao Gao: data curations, software, visualization, and writing – original draft; Long Ma: data curations, validation, and supervision; Zhe Liu: validation and supervision; Jun Huang: supervision; Hanlian Liu: supervision; Chuanzhen Huang: supervision; Yinghua Qiu: investigation, methodology, formal analysis, conceptualization, funding acquisition, and writing – review & editing.

## DATA AVAILABILITY

The data that support the findings of this study are available from the corresponding author upon reasonable request.